\PassOptionsToPackage{table}{xcolor}
\documentclass[sigconf, nonacm]{acmart}
\AtBeginDocument{%
  }

\usepackage{multirow}
\usepackage{tabularx}
\usepackage{enumitem}
\definecolor{oursrow}{gray}{0.85}
\newcommand{\sysname}{ReSID}

\setcopyright{none}
\begin{document}

\title{ReSID: Rethinking Long-Tail Retrieval in Generative Recommendation through item Reachability}

\author{Muyu Zou}
\affiliation{%
  \institution{Alibaba International Digital Commerce Group}
  \country{China}
}
\email{zoumuyu.zmy@alibaba-inc.com}
\authornote{These authors contributed equally to this research.}

\author{Shunye Tang\texorpdfstring{\textsuperscript{$\ast$}}{}}
\affiliation{%
  \institution{Nankai University}
  \country{China}
}
\email{2113298@mail.nankai.edu.cn}

\author{Yi Xu\texorpdfstring{\textsuperscript{$\ast$}}{}}
\affiliation{%
  \institution{Alibaba International Digital Commerce Group}
  \country{China}
}
\email{xy397404@alibaba-inc.com}

\author{Haibo Xing}
\affiliation{%
  \institution{Alibaba International Digital Commerce Group}
  \country{China}
}
\email{xinghaibo.xhb@alibaba-inc.com}

\author{Yandong Wang}
\affiliation{%
  \institution{Alibaba International Digital Commerce Group}
  \country{China}
}
\email{wangyandong.wyd@alibaba-inc.com}

\author{Lingyu Mu}
\affiliation{%
  \institution{Alibaba International Digital Commerce Group}
  \country{China}
}
\email{moulingyu.mly@alibaba-inc.com}

\author{Jinxin Hu}
\affiliation{%
  \institution{Alibaba International Digital Commerce Group}
  \country{China}
}
\email{jinxin.hjx@alibaba-inc.com}
\authornote{Corresponding author.}

\renewcommand{\shortauthors}{Zou et al.}

\begin{abstract}
Generative recommendation (GR) represents items with discrete Semantic IDs (SIDs) and casts retrieval as sequence generation. Under skewed interaction distributions and tight serving budgets, GR struggles to retrieve long-tail items. In this work, we are the \textbf{first} to study long-tail reachability across the complete pipeline of SID construction, decoding, and matching. We find that under equal code lengths, parallel SID construction consistently outperforms sequential SIDs on tail items. Moreover, even when full generation fails, decoded sequences carry locally consistent information that retrieves relevant tail items via partial matching. Yet this recovery is bounded by upstream search, failing once beam search prunes all prefixes completable within a fixed mismatch tolerance. Motivated by these findings, we propose ReSID to jointly preserve and recover tail candidates. Built on long parallel SIDs, ReSID pairs popularity-debiased and user-semantic-aware beam search to keep promising paths alive with partial SID matching to recover items from surviving tokens. ReSID achieves state-of-the-art long-tail retrieval while maintaining competitive overall accuracy, with lower latency and higher throughput under narrow beams. ReSID is deployed in our production system, serving main user traffic with substantial business gains. We release our complete code and model checkpoints to support full reproducibility and real-world deployment at \url{https://github.com/Gabriel-reina/ReSID}.
\end{abstract}



\maketitle

\section{Introduction}

\begin{figure}[t]
  \centering
  \includegraphics[width=\columnwidth]{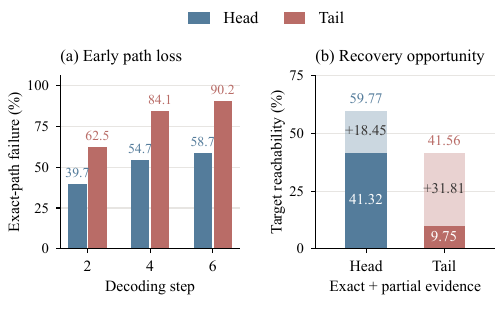}
  \caption{The long-tail reachability bottleneck on Industrial. (a) Tail targets lose their exact paths much earlier than head targets during decoding. (b) Full-SID misses still retain substantial recoverable evidence under partial matching.}
  \Description{Two empirical panels compare head and tail targets. Left: cumulative exact-prefix failure at decoding steps two, four and six. Tail failure reaches 62.49, 84.13 and 90.25 percent, versus 39.66, 54.66 and 58.68 percent for head targets. Right: exact reachability is 41.32 percent for head and 9.75 percent for tail. Partial matching adds 18.45 and 31.81 percentage points, respectively. These are candidate reachability rates, not ranked item Recall.}
  \label{fig:opening-reachability}
\end{figure}

Generative recommendation (GR) has emerged as a powerful paradigm for large-scale recommender systems~\cite{geng2023recommendationlanguageprocessingrlp,3666122.3666574,deng2025onerecunifyingretrieverank,10.1016/j.inffus.2025.103919}. Instead of scoring items directly across a massive corpus, GR represents each item as a sequence of discrete tokens, known as a Semantic ID (SID), and casts retrieval as a sequence generation task~\cite{10597986,10.1145/3624918.3625339,shi2025lladarecdiscretediffusionparallel}. Despite its effectiveness, GR struggles with the highly skewed interaction distributions typical of real-world platforms. A small fraction of popular items receives abundant supervision, whereas the vast majority of tail items suffer from sparse interactions and remain difficult to retrieve~\cite{yin2012challenginglongtailrecommendation}. This challenge is particularly acute in industrial applications, where expanding the beam size to uncover more long-tail candidates directly inflates serving costs~\cite{10.1145/3711896.3736979,10.1145/3711896.3736919,guo2026mlpsefficientdistilledgenerative}. Effective long-tail retrieval therefore demands both superior recommendation accuracy and the efficient allocation of a constrained search budget. 

Existing research on long-tail GR primarily focuses on the training stage, investigating how tail items are represented and learned. Recent studies attribute popularity bias to the under-optimization of tail SID tokens, redesigning training objectives and tokenization schemes accordingly~\cite{fan2026crabcodebookrebalancingbias,yin2026echoesfilterbubblediagnosing}. Other approaches enhance tail and cold-start generalization through adaptive SID encoding and knowledge transfer~\cite{10.1145/3805712.3808477}. While these methods improve the model's capacity to encode tail items, they treat representation learning in isolation and rely on a rigid inference pipeline of standard beam search followed by exact SID matching. Consequently, a fundamental question remains unexplored: across the complete pipeline of SID construction, decoding, and matching, how can learned tail signals survive to achieve final retrieval? In practice, this fragmented pipeline creates a severe barrier for tail items, which we term the \textit{long-tail reachability bottleneck (Figure~\ref{fig:opening-reachability})}.

\textbf{To dissect this bottleneck, we systematically study long-tail reachability across SID construction, decoding,\newline and matching, uncovering three interconnected insights (See Figure~\ref{fig:analysis-industrial-construction},~\ref{fig:analysis-industrial-recovery} and~\ref{fig:analysis-industrial-boundary}).} 
First, in SID construction, our theoretical analysis shows that sequential SIDs accumulate popularity bias across residual layers, whereas parallel SIDs quantize independent subspaces to isolate errors across positions. Under equal code lengths, parallel SIDs consistently outperform sequential SIDs on tail items across both short and long code configurations. Second, in candidate matching, because parallel SIDs isolate position-wise errors, imperfectly decoded sequences still preserve valid signals at remaining positions. Relaxing exact matching to partial matching thus provides a principled way to recover relevant tail items from these local signals.
Third, in decoding, this partial recovery is strictly bounded by upstream search. Given a fixed mismatch tolerance, recovery becomes impossible once standard beam search prematurely prunes all prefixes that can be completed within the target's tolerance neighborhood during early decoding steps.

Motivated by these findings, we propose ReSID, a reachability-aware framework that jointly preserves and recovers long-tail candidates across the retrieval pipeline (Figure~\ref{fig:model}). Built on long parallel SIDs as a representational foundation, ReSID distributes rich position-wise semantic signals across multiple tokens. To \textit{preserve} promising tail paths during decoding, ReSID introduces a popularity-debiased and user-semantic-aware beam search. Popularity debiasing prevents head items from monopolizing the narrow beam, while user behavioral semantics actively steer the search toward candidates aligned with individual user interests. Finally, to \textit{recover} relevant items during matching, ReSID pairs this guided decoding with partial SID matching, retrieving tail items from the surviving tokens.

We evaluate ReSID through extensive offline benchmarks and large-scale industrial deployment. Offline, ReSID achieves state-of-the-art performance across long-tail retrieval and coverage metrics while maintaining competitive overall recommendation accuracy. Furthermore, ReSID retrieves more relevant tail items under narrow beam constraints, directly reducing inference latency and increasing system throughput. Online, ReSID is deployed in our production system to serve main user traffic, improving ad revenue (+1.3\%), CTR (+1.1\%), GMV (+0.8\%), and long-tail exposure share (+4.0\%).

Our main contributions are summarized as follows:
\begin{itemize}
    \item \textbf{Pioneering Analysis of Long-Tail Reachability:} We are the \textbf{first} to systematically study the long-tail reachability bottleneck across the complete pipeline of SID construction, decoding, and matching. Through controlled analyses, we reveal the superiority of parallel SIDs under equal code lengths, the partial recovery opportunity from locally consistent tokens, and its upstream preservation boundary under fixed mismatch tolerance.
    \item \textbf{The ReSID Framework:} We propose ReSID, a reachability-aware generative retrieval framework that jointly preserves and recovers tail candidates. Built on long parallel SIDs, ReSID pairs popularity-debiased and user-semantic-aware beam search with partial SID matching to keep promising paths alive and recover items from surviving tokens.
    \item \textbf{Superior Effectiveness and Efficiency:} Extensive experiments demonstrate that ReSID achieves state-of-the-art long-tail retrieval and coverage while maintaining competitive overall accuracy. Its strong performance under narrow beam budgets directly translates into lower latency and higher serving throughput.
    \item \textbf{Large-Scale Industrial Deployment and Open-Source Release:} We deploy ReSID to serve main user traffic in a large-scale production recommender system, improving advertising revenue, CTR, and GMV by 1.3\%, 1.1\%, and 0.8\%, respectively, and long-tail exposure share by 4.0\%. We release our complete code and model checkpoints to support full reproducibility and real-world deployment.
\end{itemize}

\section{Related Work}

\subsection{Generative Recommendation and Semantic ID Construction}
Generative recommendation (GR) recasts retrieval as autoregressive generation over discrete Semantic IDs (SIDs). Pure RQ-VAE quantizes item content vectors into hierarchical codes while retaining one-to-many collisions~\cite{9879532}. TIGER resolves such collisions by appending a disambiguation token, whereas OneRec uses residual $k$-means with capacity balancing to constrain item allocation per code~\cite{3666122.3666574,deng2025onerecunifyingretrieverank}. However, these methods pair sequential residual SIDs with standard beam search and exact full-SID matching, overlooking how SID structure interacts with inference. In contrast, we show that under equal code lengths, parallel SIDs consistently outperform sequential SIDs on tail items, offering strong position-wise signals for fault-tolerant retrieval.

\subsection{Long-Tail Modeling in Generative Recommendation}
In GR, skewed interactions cause popular items to dominate shared prefixes during decoding. Recent studies address this via training-side tokenization and objectives. CRAB splits popular codes under tree constraints and adaptive training~\cite{fan2026crabcodebookrebalancingbias}. Ghost lets tail items inherit head prefixes with extended suffixes under unlikelihood training, while SA$^2$CRQ combines adaptive code lengths with head-anchored tail expansion~\cite{yin2026echoesfilterbubblediagnosing,10.1145/3805712.3808477}. Yet these methods leave inference untouched, where beam search prunes tail paths and exact matching discards imperfectly decoded sequences. Moreover, as noted in Ghost, certain training-side components improve tail retrieval at the expense of overall accuracy. To our knowledge, ReSID is the \textbf{first} framework to address long-tail reachability across SID construction, decoding, and matching, with every component boosting tail retrieval without compromising overall performance.

\section{Empirical Analysis of Long-Tail Reachability}
\label{sec:analysis}

In generative recommendation, retrieving an item from the catalog spans three stages, namely offline Semantic ID construction, online beam search decoding, and downstream candidate matching. Existing studies on long-tail generative recommendation primarily focus on improving tokenization or training objectives during the training stage while keeping the inference pipeline fixed. Before introducing our proposed method, we first conduct controlled analyses across all three stages in this section to investigate where long-tail items are lost along the retrieval pipeline and motivate our framework design.

\subsection{Preliminaries and Experimental Setup}
In standard generative recommendation, an offline tokenizer maps each item into a fixed-length sequence of discrete tokens known as a Semantic ID~\cite{10.1145/3627673.3679569,10.1145/3640457.3688178,zheng2025universalitemtokenizationtransferable}. During online serving, an autoregressive model predicts next-token probabilities step by step from the user history, while beam search retains the most probable prefix paths at each step until a set of complete token sequences is generated~\cite{2969033.2969173}. Finally, the system performs exact matching and returns only those items whose complete codes appear in the generated set.

We conduct our primary analyses on the Industrial dataset using a standard generative recommendation backbone under a tight beam budget. Target items are grouped into head and tail categories according to their training frequencies and evaluated using accuracy and coverage metrics. Metric definitions are given in Section~\ref{sec:exp-setup}, and dataset statistics and parallel analyses on the Office dataset are provided in the Appendix.

\subsection{Analysis of the Reachability Bottleneck}
Through controlled experiments across SID construction, candidate matching, and beam search decoding, we uncover three key findings regarding how offline code structure, local token overlap, and early prefix pruning jointly govern long-tail reachability. Below, we examine each finding in detail.

\begin{figure}[t]
  \centering
  \includegraphics[width=\columnwidth]{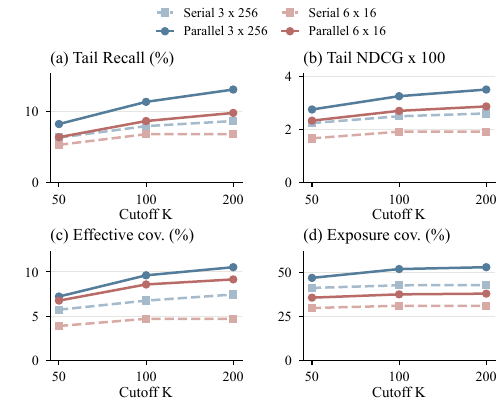}
  \caption{Effect of SID construction on tail retrieval (Industrial; beam 20, exact matching). The nominal code space is identical: $256^3=16^6$. Dashed squares: serial; solid circles: parallel. All four metrics are shown at $K=50,100,200$.}
  \Description{Four line charts compare tail Recall, NDCG, effective coverage, and exposure coverage for serial and parallel SID construction with 3 by 256 and 6 by 16 codebooks.}
  \label{fig:analysis-industrial-construction}
\end{figure}

\textbf{Under equal code lengths, parallel SIDs are more favorable for long-tail retrieval than sequential SIDs.} We first examine how offline SID construction affects tail retrieval (Figure~\ref{fig:analysis-industrial-construction}). Mainstream sequential construction quantizes the residual of each preceding layer hierarchically, whereas parallel construction partitions item representations into distinct subspaces and quantizes each subspace independently. Holding the nominal code space identical ($256^3 = 16^6$), we compare sequential and parallel SIDs under both short ($3\times 256$) and long ($6\times 16$) code configurations. Across both settings, parallel SIDs consistently outperform sequential SIDs on all tail accuracy and coverage metrics, raising tail Recall@200 from 8.62\% to 13.02\% under the short configuration and nearly doubling tail effective coverage from 4.70\% to 9.16\% under the long configuration. In sequential SIDs, each token depends on the residual of preceding layers; because head items already dominate early prefixes, passing dependencies layer by layer further amplifies this head dominance at subsequent positions. 
In contrast, parallel construction decouples semantic subspaces across positions to prevent such cumulative bias. Appendix~\ref{app:theory} theoretically analyzes this residual-bias accumulation and its isolation by parallel construction.

\begin{figure}[t]
  \centering
  \includegraphics[width=\columnwidth]{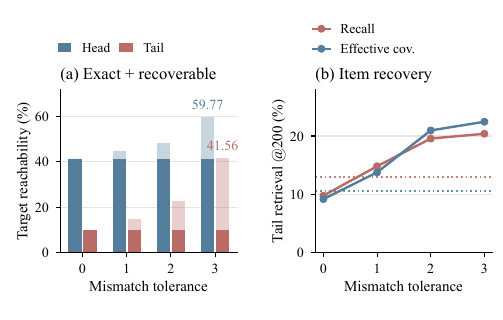}
  \caption{Local evidence and item recovery (Industrial; fixed beam 20). (a) Fraction of targets within the mismatch tolerance of a single complete beam: solid segments denote exact hits; pale segments show additional recoverable targets. (b) Tail Recall and effective coverage at $K=200$ after partial matching of parallel $6\times16$ SIDs. Dotted lines: parallel $3\times256$ exact matching. Panel (a) measures candidate reachability; panel (b) measures ranked item retrieval.}
  \Description{Left: stacked head and tail reachability bars separate exact hits and additional recoverable targets for zero to three mismatches. Right: tail Recall and effective coverage after matching, compared with the short parallel exact-match baseline.}
  \label{fig:analysis-industrial-recovery}
\end{figure}

\textbf{Even when full generation fails, decoded sequences retain locally consistent signals that recover tail items via partial matching.} 
Having examined the impact of SID construction, we next turn to SID matching. We find that despite the benefits of parallel construction, tail retrieval under exact matching still lags far behind head retrieval, with longer codes performing even worse than shorter ones because requiring every single position to be correct is far more stringent. However, as established in our theoretical analysis above, parallel SIDs decouple semantic subspaces across positions to isolate errors. Consequently, even when full-sequence generation fails, the decoded sequences still preserve valid item signals at the remaining positions. Figure~\ref{fig:analysis-industrial-recovery}(a) directly confirms this property. While requiring zero mismatches leaves a massive gap between head and tail reachability (41.32\% vs.\ 9.75\%), tolerating one to three mismatched positions boosts tail reachability more than fourfold to 41.56\%, substantially narrowing the gap with head items (59.77\%). In fact, over one-third of missed tail targets already share at least half of their tokens with a single decoded sequence. Relaxing the matching requirement to allow partial matches over the same decoded beams (Figure~\ref{fig:analysis-industrial-recovery}(b)) therefore more than doubles both tail recall@200 (from 9.75\% to 20.39\%) and effective coverage (from 9.16\% to 22.45\%) of the long parallel model, surpassing the short parallel baseline.

\begin{figure}[t]
  \centering
  \includegraphics[width=\columnwidth]{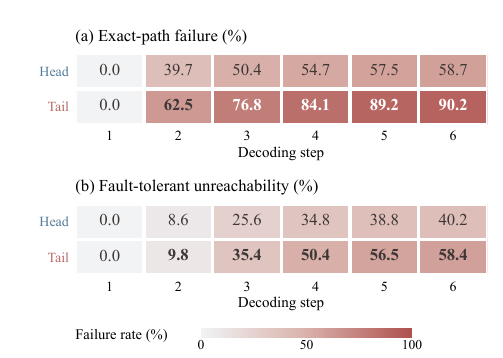}
  \caption{Early beam-search failure on Industrial (parallel $6\times16$, beam 20). (a) Exact-path failure: the target prefix has left the beam. (b) Fault-tolerant unreachability: no retained prefix admits a catalog-valid completion within three mismatches of the target. Darker cells indicate higher failure. Final exact failure is 90.25\% for tail versus 58.68\% for head; 50.36\% of tail targets are already unrecoverable by step 4.}
  \Description{Two annotated heatmaps show head and tail cumulative failures across all six decoding steps. The upper heatmap measures exact prefix loss; the lower measures loss of all catalog-valid completions within three mismatches. Tail failure is substantially higher and occurs early.}
  \label{fig:analysis-industrial-boundary}
\end{figure}

\textbf{Early prefix pruning during upstream beam search bounds the recovery ceiling of downstream partial matching.} Relaxing the matching constraint alone is still not enough. Even when we allow up to three mismatched tokens, 58.4\% of tail targets remain unreachable, and relaxing the tolerance further brings diminishing returns while introducing irrelevant noise. We therefore turn our attention upstream to beam search decoding (Figure~\ref{fig:analysis-industrial-boundary}) to investigate why so many tail items fail to survive to the matching stage.

We first track the loss of exact target prefixes at each decoding step in Figure~\ref{fig:analysis-industrial-boundary}(a). After generating just two tokens, 62.5\% of tail targets have already fallen out of the beam. By the final step, tail failure reaches a staggering 90.2\%, whereas head failure is much lower at 58.7\%. Next, to evaluate the upper bound of downstream partial matching during decoding, we relax this exact requirement in Figure~\ref{fig:analysis-industrial-boundary}(b). Specifically, we check whether the beam retains any prefix that can be completed into a valid catalog SID within three mismatches of the target. Even under this relaxed criterion, the results remain poor. By step 4, while most head targets (65.2\%) still keep at least one recoverable prefix in the beam, over half of the tail targets (50.4\%) have already lost every recoverable path, and this failure rate rises to 58.4\% by the final step. This analysis demonstrates that recovering tail items solely at the matching stage is insufficient, because a large fraction of tail items becomes completely unrecoverable before beam search even finishes. Consequently, beam search itself must be modified to help promising tail prefixes survive pruning.

Taken together, our three stage-wise analyses reveal a clear roadmap for long-tail retrieval across the complete pipeline. First, parallel SID construction decouples token positions to prevent head dominance from accumulating across layers, providing reliable position-wise signals. Second, partial SID matching leverages these local signals to recover relevant tail items when full generation fails. Third, guided beam search is essential to prevent promising tail prefixes from being pruned prematurely so that they can reach the matching stage. These three findings directly motivate our framework design in the next section.

\begin{figure*}[t]
  \centering
  \includegraphics[width=\textwidth]{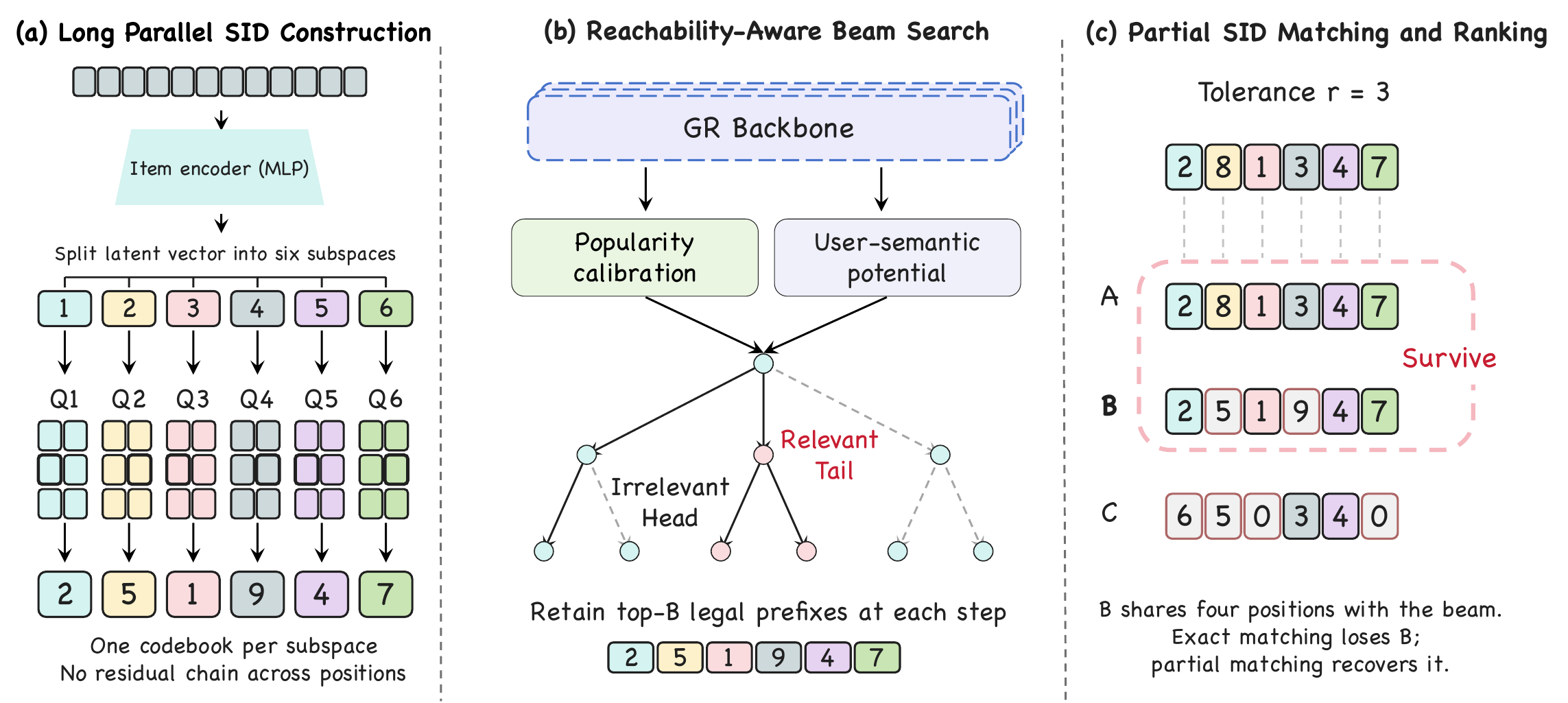}
  \caption{Overview of the ReSID framework. (a)~\textbf{Long Parallel SID Construction}: Item representations are partitioned into distinct subspaces and quantized by independent codebooks without cross-position residual chains. (b)~\textbf{Reachability-Aware Beam Search}: Popularity calibration and user-semantic potential jointly guide step-wise prefix expansion to retain promising tail branches within the top-$B$ legal prefixes. (c)~\textbf{Partial SID Matching and Ranking}: Rather than requiring exact sequence matches, partial matching recovers candidate items that share sufficient position-wise tokens with the decoded beams.}
  \Description{Schematic diagram of the ReSID framework with three components: (a) Long Parallel SID Construction, which splits item latent vectors into independent subspaces quantized by separate codebooks; (b) Reachability-Aware Beam Search, which combines popularity calibration and user-semantic potential to retain relevant tail prefixes during decoding; and (c) Partial SID Matching and Ranking, which recovers partially matched items within the allowed mismatch tolerance.}
  \label{fig:model}
\end{figure*}

\section{The ReSID Framework}
\label{sec:resid}

Based on Section~\ref{sec:analysis}, we propose ReSID to improve long-tail reachability across the entire retrieval pipeline (Figure~\ref{fig:model}). ReSID has three stages. First, it builds long parallel SIDs so that each token position captures an independent semantic aspect. Second, it guides beam search with popularity debiasing and user-semantic scoring to keep relevant tail prefixes in the beam. Third, it uses partial SID matching to recover tail items from the decoded sequences.

\subsection{Long Parallel SID Construction}
\label{sec:lpsid}
Sequential SIDs propagate upstream quantization shifts through their residual inputs (Appendix~\ref{app:theory}). ReSID isolates this propagation by quantizing separate latent subspaces in parallel~\cite{10.1109/TPAMI.2010.57,10.1145/3711896.3736979}.

Given an item text embedding $\mathbf{x}_i$, an MLP encoder maps it into a latent vector $\mathbf{z}_i$ and splits $\mathbf{z}_i$ evenly into $L$ subspaces $\mathbf{z}_i = \mathbf{z}_i^{(1)} \Vert \cdots \Vert \mathbf{z}_i^{(L)}$. Each position $\ell$ has its own codebook $E_\ell = \allowbreak\{\mathbf{e}_{\ell,0}, \allowbreak\dots, \allowbreak\mathbf{e}_{\ell,K-1}\}$ and assigns $\mathbf{z}_i^{(\ell)}$ to the nearest codeword:
\begin{equation}
\begin{aligned}
c_{i,\ell} &= \arg\min_{k} \| \mathbf{z}_i^{(\ell)} - \mathbf{e}_{\ell,k} \|_2^2, \\
\hat{\mathbf{z}}_i &= \mathbf{e}_{1,c_{i,1}} \Vert \cdots \Vert \mathbf{e}_{L,c_{i,L}}.
\end{aligned}
\end{equation}
A decoder reconstructs $\mathbf{x}_i$ from $\hat{\mathbf{z}}_i$ under standard reconstruction and commitment losses~\cite{3295222.3295378}. Each item is then represented by a parallel SID $\mathbf{c}_i = (c_{i,1}, \dots, c_{i,L})$. On top of these codes, we train a standard Transformer recommender~\cite{vaswani2023attentionneed} with cross-entropy loss to encode user history $H_u$ and predict next-token probabilities $P_\theta(c_{i,\ell} \mid H_u, c_{i,<\ell})$.

\subsection{Reachability-Aware Guided Beam Search}
\label{sec:rabs}
Partial matching can recover a target item only if at least one sequence within $r$ mismatches survives beam search. Under a small beam size $B$, however, standard beam search prunes most tail prefixes early on. We therefore freeze the base recommender and add two guidance signals during beam search.

\textbf{Popularity Calibration.} In standard decoding, popular items appear frequently in training data and receive much higher transition probabilities. As a result, head prefixes quickly fill up the narrow beam and push tail prefixes out. To reduce this bias, we estimate token transition priors from training frequencies and subtract the scaled log-prior from the model logits before softmax~\cite{menon2021longtail}. This penalty lowers the scores of overly popular branches and leaves room in the beam for tail prefixes.

\textbf{User-Semantic Alignment.} Debiasing popularity alone is not enough. If we only penalize popular branches, the beam will keep random tail prefixes that do not match the user's interest. We therefore add a user-semantic score to identify which tail branches are actually relevant to the user. Specifically, a linear layer maps the pooled encoder output into a normalized user vector, which is trained with cross-entropy loss to align with the normalized text vector of the target item~\cite{oord2019representationlearningcontrastivepredictive}. During decoding, we measure how well a prefix matches the user's interest by averaging the user--item similarity over all valid catalog items under that prefix:
\begin{equation}
\Phi_u(p) = \log \left[ \frac{1}{|\mathcal{I}(p)|} \sum_{i \in \mathcal{I}(p)} \exp\left( \frac{\mathbf{q}_u^\top \mathbf{t}_i}{\tau} \right) \right],
\end{equation}
where $\mathcal{I}(p)$ is the valid item set under prefix $p$, $\mathbf{q}_u$ and $\mathbf{t}_i$ are normalized user and item vectors, and $\tau$ is the temperature.

Combining both signals, we score each expanded candidate branch at each decoding step as:
\begin{equation}
S_u(p \Vert v) = S_u(p) + \log \tilde{P}_\theta(v \mid H_u, p) + \beta \big[ \Phi_u(p \Vert v) - \Phi_u(p) \big],
\end{equation}
where $p \Vert v$ denotes prefix $p$ extended by token $v$, $\tilde{P}_\theta(v \mid H_u, p)$ is the popularity-calibrated probability obtained by subtracting $\lambda \log \pi(v \mid p)$ from the logits under empirical prior $\pi(v \mid p)$, $\Phi_u(p \Vert v) - \Phi_u(p)$ measures how much the new branch improves user relevance, and $\lambda, \beta \ge 0$ weight the two terms. At each step, beam search keeps the top-$B$ valid prefixes~\cite{decao2021autoregressiveentityretrieval,10.1145/3770855.3818506} under this score, focusing the beam on tail prefixes that are both less popular and relevant to the user. These semantic scores can be cached and reused across requests with unchanged user states via prefix-level lookups.


\subsection{Partial SID Matching and Candidate Ranking}
\label{sec:psm}
After $L$ decoding steps, beam search returns $B$ complete SID sequences. Standard generative retrieval relies on exact matching and discards any decoded sequence that does not fully match a catalog item. However, as shown in Section~\ref{sec:analysis}, because parallel SIDs isolate errors across independent positions, decoded sequences still retain locally consistent token signals that point to relevant tail items even when full-sequence generation fails.

Motivated by this property, we replace exact matching with partial SID matching under a mismatch tolerance $r$. Specifically, we compare catalog items against the $B$ decoded sequences and retrieve all candidate items whose SIDs differ from at least one decoded sequence by at most $r$ tokens. Finally, we rank the retrieved candidates by prioritizing those with fewer mismatched tokens. For candidates with the same number of mismatches, those matched by higher-scoring beams are ranked higher.

\begin{table*}[t]
\caption{Performance comparison between \sysname{} and baselines across three datasets, on all requests (All) and on long-tail requests (Tail). Recall and coverage are in \%. The best results are in bold and the second-best results are underlined.}
\label{tab:main}
\centering\scriptsize
\setlength{\tabcolsep}{6.55pt}
\begin{tabular}{@{}llcccccccccccccccc@{}}
\toprule
\multirow{2}{*}{Dataset} & \multirow{2}{*}{Method} & \multicolumn{2}{c}{R@50} & \multicolumn{2}{c}{R@100} & \multicolumn{2}{c}{R@200} & \multicolumn{2}{c}{N@50} & \multicolumn{2}{c}{N@100} & \multicolumn{2}{c}{N@200} & \multicolumn{2}{c}{EffCov@200} & \multicolumn{2}{c}{ExpCov@200} \\
\cmidrule(lr){3-4} \cmidrule(lr){5-6} \cmidrule(lr){7-8} \cmidrule(lr){9-10} \cmidrule(lr){11-12} \cmidrule(lr){13-14} \cmidrule(lr){15-16} \cmidrule(lr){17-18}
 & & All & Tail & All & Tail & All & Tail & All & Tail & All & Tail & All & Tail & All & Tail & All & Tail \\
\midrule
\multirow{7}{*}{Industrial} & RQ-VAE & 18.20 & 3.27 & 25.79 & 5.05 & 34.24 & 7.79 & 0.0881 & 0.0115 & 0.1004 & 0.0144 & 0.1123 & 0.0182 & 25.16 & 6.99 & 85.08 & 74.52 \\
 & TIGER & 20.10 & 1.01 & 25.79 & 1.96 & 33.09 & 4.58 & \textbf{0.1080} & 0.0028 & \textbf{0.1172} & 0.0043 & \textbf{0.1274} & 0.0079 & 24.93 & 5.50 & 87.30 & 78.71 \\
 & OneRec & \textbf{22.48} & 6.78 & \textbf{28.52} & 9.81 & \textbf{36.66} & 14.68 & \underline{0.1061} & 0.0194 & \underline{0.1159} & 0.0243 & \underline{0.1273} & 0.0310 & 30.60 & 14.43 & 85.73 & 75.93 \\
\cmidrule(l){2-18}
 & CRAB & 19.72 & 0.89 & 24.99 & 1.84 & 32.03 & 4.70 & 0.1060 & 0.0023 & 0.1144 & 0.0039 & 0.1243 & 0.0078 & 24.99 & 6.30 & 88.09 & 80.58 \\
 & Ghost & 19.26 & \underline{7.31} & 23.54 & \underline{9.93} & 31.63 & \underline{16.11} & 0.1034 & \underline{0.0211} & 0.1104 & \underline{0.0254} & 0.1216 & \underline{0.0339} & \underline{32.03} & \underline{21.42} & \underline{94.90} & \underline{92.84} \\
 & SA$^2$CRQ & 20.41 & 2.91 & 27.18 & 4.82 & \underline{35.54} & 7.31 & 0.0882 & 0.0092 & 0.0992 & 0.0123 & 0.1108 & 0.0158 & 24.76 & 6.07 & 50.05 & 29.97 \\
\cmidrule(l){2-18}
\rowcolor{oursrow} & \textbf{\sysname{}} & \underline{22.02} & \textbf{9.22} & \underline{27.47} & \textbf{14.03} & 33.77 & \textbf{21.22} & 0.1025 & \textbf{0.0297} & 0.1113 & \textbf{0.0375} & 0.1201 & \textbf{0.0476} & \textbf{33.01} & \textbf{25.20} & \textbf{99.40} & \textbf{99.09} \\
\addlinespace
\multirow{7}{*}{Office} & RQ-VAE & 21.13 & 1.92 & 26.96 & 2.63 & 34.18 & 4.34 & 0.1174 & 0.0064 & 0.1269 & 0.0075 & 0.1370 & 0.0099 & 27.64 & 6.17 & 89.48 & 81.81 \\
 & TIGER & 21.74 & 1.56 & 27.15 & 2.20 & 34.03 & 4.48 & \textbf{0.1233} & 0.0065 & \underline{0.1320} & 0.0075 & \underline{0.1416} & 0.0106 & 28.20 & 5.84 & 91.21 & 84.85 \\
 & OneRec & \textbf{22.96} & 3.20 & \textbf{28.55} & 5.12 & \textbf{36.42} & 8.32 & \underline{0.1232} & 0.0155 & \textbf{0.1322} & 0.0186 & \textbf{0.1432} & 0.0230 & 31.99 & 10.25 & 88.44 & 79.73 \\
\cmidrule(l){2-18}
 & CRAB & 20.65 & 1.85 & 26.26 & 2.70 & 32.41 & 4.83 & 0.1199 & 0.0075 & 0.1290 & 0.0088 & 0.1376 & 0.0118 & 28.30 & 6.28 & 92.31 & 86.99 \\
 & Ghost & 18.17 & \underline{5.90} & 22.52 & \underline{8.60} & 30.44 & \underline{13.50} & 0.1158 & \underline{0.0196} & 0.1228 & \underline{0.0240} & 0.1338 & \underline{0.0308} & \underline{34.54} & \underline{17.75} & \underline{96.62} & \underline{94.83} \\
 & SA$^2$CRQ & 21.37 & 1.28 & \underline{27.17} & 2.06 & \underline{35.12} & 3.41 & 0.1109 & 0.0039 & 0.1203 & 0.0052 & 0.1313 & 0.0071 & 23.95 & 4.19 & 45.33 & 26.88 \\
\cmidrule(l){2-18}
\rowcolor{oursrow} & \textbf{\sysname{}} & \underline{21.80} & \textbf{8.10} & 26.24 & \textbf{13.65} & 31.52 & \textbf{20.54} & 0.1159 & \textbf{0.0234} & 0.1232 & \textbf{0.0325} & 0.1306 & \textbf{0.0421} & \textbf{36.28} & \textbf{25.03} & \textbf{99.54} & \textbf{99.31} \\
\addlinespace
\multirow{7}{*}{In-house} & RQ-VAE & 28.42 & 8.38 & 33.01 & 9.84 & 37.59 & 11.32 & 0.1416 & 0.0411 & 0.1490 & 0.0439 & 0.1554 & 0.0460 & 40.42 & 11.47 & 52.51 & 43.89 \\
 & TIGER & \underline{28.56} & 7.22 & \underline{33.12} & 8.47 & \underline{37.65} & 9.72 & \underline{0.1434} & 0.0361 & \underline{0.1508} & 0.0381 & \underline{0.1572} & 0.0399 & \underline{40.47} & 9.84 & 52.52 & 42.21 \\
 & OneRec & 28.11 & 8.28 & 32.72 & 9.75 & 37.38 & 11.27 & 0.1380 & 0.0404 & 0.1455 & 0.0428 & 0.1520 & 0.0449 & 40.21 & 11.41 & 52.53 & 42.09 \\
\cmidrule(l){2-18}
 & CRAB & 23.67 & 5.25 & 27.94 & 6.17 & 32.39 & 7.17 & 0.1135 & 0.0258 & 0.1204 & 0.0273 & 0.1267 & 0.0287 & 35.06 & 7.27 & 51.88 & 33.40 \\
 & Ghost & 23.28 & \underline{13.77} & 27.54 & \underline{16.24} & 32.07 & \underline{18.94} & 0.1092 & \underline{0.0661} & 0.1161 & \underline{0.0701} & 0.1224 & \underline{0.0734} & 34.73 & \underline{19.23} & 51.77 & \underline{49.08} \\
 & SA$^2$CRQ & 25.94 & 10.45 & 30.61 & 12.31 & 35.56 & 14.30 & 0.1232 & 0.0509 & 0.1309 & 0.0539 & 0.1378 & 0.0567 & 38.48 & 14.51 & \underline{57.00} & 24.56 \\
\cmidrule(l){2-18}
\rowcolor{oursrow} & \textbf{\sysname{}} & \textbf{35.81} & \textbf{26.59} & \textbf{42.02} & \textbf{31.56} & \textbf{48.28} & \textbf{36.62} & \textbf{0.1777} & \textbf{0.1313} & \textbf{0.1878} & \textbf{0.1394} & \textbf{0.1967} & \textbf{0.1464} & \textbf{54.65} & \textbf{38.80} & \textbf{89.94} & \textbf{73.11} \\
\bottomrule
\end{tabular}
\end{table*}

\section{Experiments}
In this section, we evaluate the effectiveness and efficiency of
\sysname{} on two public benchmarks and a large-scale industrial dataset.

\subsection{Experimental Setup}
\label{sec:exp-setup}


\subsubsection{Datasets and Evaluation Metrics.}
We conduct experiments on two categories of the Amazon Reviews dataset~\cite{hou-etal-2026-bridging}, \emph{Industrial} and \emph{Office}, and on a large-scale \emph{In-house} dataset collected from a leading e-commerce advertising platform. On the two public benchmarks, we define \emph{long-tail} items as those appearing 1--6 times as training targets and \emph{head} items as those appearing at least 12 times. On the In-house dataset, we rank items by their training click frequencies and define the smallest leading set accounting for at least 80\% of clicks as the head, with the remaining observed items forming the long tail. All partitions are determined from the training split and fixed during evaluation, and dataset statistics are given in Appendix~\ref{app:datasets}. We adopt Recall and NDCG at $K\in\{50,100,200\}$~\cite{10.1145/582415.582418}, denoted by R@$K$ and N@$K$, together with two coverage metrics at $K{=}200$: \emph{effective coverage} (EffCov), the fraction of distinct target items that are correctly retrieved at least once, and \emph{exposure coverage} (ExpCov), the fraction of catalog items that appear in any retrieved list. All metrics are reported on all test requests (\emph{All}) and on the requests whose target is a long-tail item (\emph{Tail}). Results at $K\in\{5,10,20\}$ are reported in Appendix~\ref{app:smallk}.


\subsubsection{Baselines.}
We compare \sysname{} with two groups of baselines. The first group consists
of representative SID-based GR methods, \emph{i.e.}, RQ-VAE~\cite{9879532},
TIGER~\cite{3666122.3666574} and OneRec~\cite{deng2025onerecunifyingretrieverank}; the second group consists of recent methods designed for long-tail
generative retrieval, \emph{i.e.}, CRAB~\cite{fan2026crabcodebookrebalancingbias}, Ghost~\cite{yin2026echoesfilterbubblediagnosing}
and SA$^2$CRQ~\cite{10.1145/3805712.3808477}. Each baseline is briefly described in
Appendix~\ref{app:baselines}. For a fair comparison, all methods are
implemented on the same lightweight encoder--decoder GR backbone and differ
only in their SID scheme or core long-tail mechanism.

\subsubsection{Implementation Details.}
On the public benchmarks, \sysname{} uses parallel SIDs with $L{=}6$ positions and 16 codes per position, decoding with a beam width of 20 whereas all baselines use a beam width of 200. On the In-house dataset, to accommodate its much larger item catalog and richer semantics, we scale the SIDs to $L{=}8$ positions and 300 codes per position, with a beam width of 100 for \sysname{} and 300 for all baselines. We set the popularity-calibration strength to $\lambda{=}0.5$ and the user-semantic weight to $\beta{=}0.4$, and Partial SID Matching allows at most $r{=}3$ mismatches. These guidance and matching hyper-parameters are kept identical across all datasets without per-dataset tuning, and their sensitivity is analyzed in Section~\ref{sec:exp-sens}. Further details are provided in Appendix~\ref{app:impl}.


\subsection{Overall Performance}
\label{sec:exp-main}

Table~\ref{tab:main} reports the performance comparison on the three
datasets.

\sysname{} achieves the best results on all long-tail metrics across the
three datasets. Compared with Ghost, the strongest baseline on every
long-tail column, it improves long-tail Recall and NDCG by 41.2\% and
36.6\% on average over the three cutoffs and the two public datasets.
On In-house, tail R@200 rises from 18.94 to 36.62 and N@200 from
0.0734 to 0.1464. These gains come with a broader reach: \sysname{}
attains the highest long-tail EffCov@200 on all datasets, while ExpCov@200
exceeds 99\% on the public datasets and reaches 73.11\% on In-house,
showing that the additional exposure translates into correct retrievals.

On all requests, \sysname{} remains competitive with the strongest baselines.
It ranks second on overall R@50 on both public datasets, and its average
gap to the best result per metric is 6.7\% in Recall and 6.2\% in NDCG.
On In-house, it leads on every metric, raising overall R@200 from the
strongest baseline's 37.65 to 48.28. The lead of the representative GR
methods on overall metrics on the public datasets is mainly a consequence
of their skewed popularity. As shown in Table~\ref{tab:stats}, long-tail
items make up more than half of the catalog but receive only 20--22\%
of the training interactions, while head items receive 56--63\%.
These methods mainly serve head targets; on Office, for example, OneRec
reaches an overall R@200 of 36.42 but only 8.32 on long-tail requests,
in line with the head--tail gap of a standard GR model analyzed in
Section~\ref{sec:analysis}.
In contrast, \sysname{} obtains the highest overall EffCov and ExpCov,
indicating that it reaches more items instead of concentrating on popular ones.

Moreover, \sysname{} achieves these results with beam widths of 20 versus
200 on the public datasets and 100 versus 300 on In-house, suggesting that
its gains stem from preserving and recovering relevant candidates rather
than from a larger search budget. The resulting serving-cost advantage
is quantified in Section~\ref{sec:exp-eff}.

\subsection{Ablation Study}
\label{sec:exp-ablation}

\begin{table}[t]
\caption{Ablation study of \sysname{}. Coverage is measured at $K{=}200$, and $\Delta$ denotes the relative change in R@200 with respect to \sysname{}. The best results in each group are in bold.}
\label{tab:ablation}
\centering\scriptsize
\setlength{\tabcolsep}{1.5pt}
\begin{tabular}{llccccccc}
\toprule
 & Variant & R@50 & R@100 & R@200 & $\Delta$(\%) & N@200 & EffCov & ExpCov \\
\midrule
\rowcolor{oursrow} & \sysname{} & 21.80 & \textbf{26.24} & \textbf{31.52} & -- & \textbf{0.1306} & \textbf{36.28} & 99.54 \\
 & \quad w/o popularity calibration & 21.19 & 26.08 & 31.30 & -0.7 & 0.1287 & 34.19 & 98.64 \\
 & \quad w/o user-semantic alignment & 20.16 & 24.97 & 30.05 & -4.7 & 0.1259 & 34.54 & 99.77 \\
 & \quad w/o partial matching & 20.80 & 21.13 & 21.13 & -33.0 & 0.1147 & 20.42 & 77.74 \\
\multirow{-5}{*}{\rotatebox{90}{All}} & \quad w/ sequential SIDs & \textbf{22.13} & 25.05 & 29.47 & -6.5 & \textbf{0.1306} & 33.83 & \textbf{99.86} \\
\addlinespace
\rowcolor{oursrow} & \sysname{} & \textbf{8.10} & \textbf{13.65} & \textbf{20.54} & -- & \textbf{0.0421} & \textbf{25.03} & 99.31 \\
 & \quad w/o popularity calibration & 6.47 & 12.58 & 18.76 & -8.7 & 0.0378 & 22.71 & 97.81 \\
 & \quad w/o user-semantic alignment & 7.46 & 13.57 & 19.26 & -6.2 & 0.0390 & 23.81 & 99.63 \\
 & \quad w/o partial matching & 6.82 & 6.97 & 6.97 & -66.1 & 0.0213 & 8.60 & 63.73 \\
\multirow{-5}{*}{\rotatebox{90}{Tail}} & \quad w/ sequential SIDs & 7.68 & 11.37 & 16.99 & -17.3 & 0.0338 & 21.39 & \textbf{99.79} \\
\bottomrule
\end{tabular}
\end{table}

To investigate the contribution of each component, we conduct ablation
studies. Starting from the full model, we remove popularity
calibration (Section~\ref{sec:rabs}) by setting $\lambda{=}0$, remove
user-semantic alignment by setting $\beta{=}0$, and remove partial matching
(Section~\ref{sec:psm}) by allowing exact matches only; we further replace the
parallel SIDs with sequential SIDs of the same length. Based on
Table~\ref{tab:ablation}, we have the following observations.
(1)~\emph{Partial matching is the primary source of gains.} Removing it
reduces R@200 by 33.0\% on all requests and by 66.1\% on long-tail requests,
and lowers long-tail ExpCov@200 from 99.31\% to 63.73\%, confirming that
exact matching discards a large amount of useful SID evidence.
(2)~\emph{Both signals of Reachability-Aware Guided Beam Search are beneficial.}
Removing popularity calibration or user-semantic alignment reduces long-tail R@200 by
8.7\% and 6.2\%, respectively, and also lowers overall R@200 and EffCov@200,
indicating that preserving tail candidates during search does not come at the
expense of overall quality. (3)~\emph{Parallel SIDs improve long-tail
retrieval without compromising overall performance.} Replacing them with
sequential SIDs reduces long-tail R@200 by 17.3\% and long-tail EffCov@200 by
3.64 points, and also lowers overall R@200 by 6.5\% and overall EffCov@200 by
2.45 points, while overall R@50 and NDCG remain comparable.

\subsection{Fine-Grained Mechanism Analysis}
\label{sec:mech}
While the ablation study measures the overall gain of each component, we further conduct four fine-grained analyses on Industrial and Office to examine how ReSID works at the request and branch levels. Figure~\ref{fig:mech-summary} summarizes the main results at $K=200$, and complete tables are provided in Appendix~\ref{app:mech}.

\begin{figure}[!htbp]
\centering
\includegraphics[width=\columnwidth]{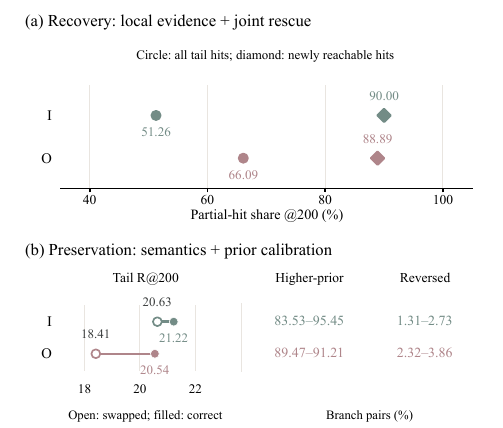}
\caption{Four mechanism analyses in two panels (I: Industrial; O: Office). (a) Partial matching's share of all tail hits and newly reachable tail hits at $K=200$. (b) Correct versus swapped user semantics (left); higher-prior rivals and pairwise rank reversals under calibration (right, ranges over steps 4--6). Full details are in Appendix~\ref{app:mech}.}
\Description{Recovery panel summarizes partial matching and joint rescue; preservation panel summarizes semantic specificity and fixed-pool popularity calibration.}
\label{fig:mech-summary}
\end{figure}

\begin{figure*}[t]
  \centering
  \includegraphics[width=\textwidth]{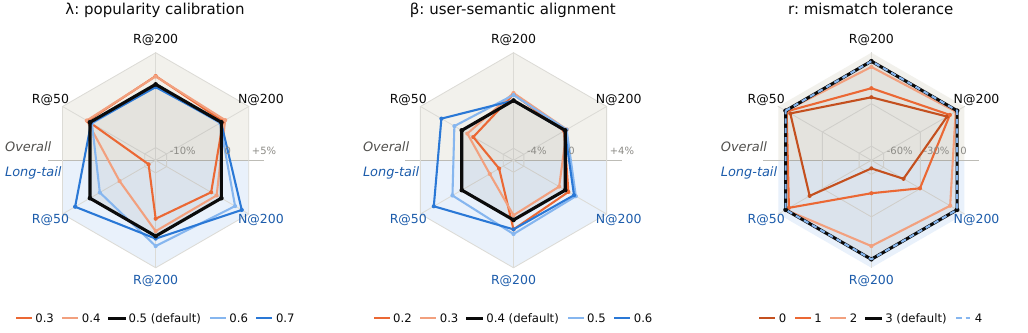}
  \caption{Hyper-parameter analysis of \sysname{} on the Office validation
  set. Each spoke shows a metric relative to the default configuration
  ($\lambda{=}0.5$, $\beta{=}0.4$, $r{=}3$, in black), with all requests in the
  upper half and long-tail requests in the lower half. Values below and above
  the default are drawn in orange and blue, respectively.}
  \Description{Three radar charts compare metrics with the default configuration on the Office validation set. Popularity calibration ranges from 0.3 to 0.7, user-semantic weight from 0.2 to 0.6, and mismatch tolerance from 0 to 4. Each chart has six spokes, R@50, N@200 and R@200, for all requests in the upper half and long-tail requests in the lower half. Curves for the mismatch tolerance spread widest, whereas the calibration and semantic-weight curves stay close to the default.}
  \label{fig:sensitivity}
\end{figure*}

\textbf{Partial matching accounts for the majority of tail hits.} We first examine how many tail hits in full ReSID have no exact match in the decoded beams and rely strictly on partial matching. At $K=200$, partial matching contributes 51.26\% of all tail hits (183 out of 357) on Industrial and 66.09\% (191 out of 289) on Office. Thus, more than half of the retrieved tail items come from locally consistent tokens rather than exact full-SID generation.

\textbf{Guided search and partial matching jointly rescue unreachable items.} Next, we examine tail targets unreachable within three mismatches under standard beam search, yet retrieved by ReSID at $K=200$. On Industrial and Office, 90.00\% (9 out of 10) and 88.89\% (8 out of 9) of these newly rescued hits still rely on partial matching after guided search preserves their fault-tolerant prefixes. This confirms that guided decoding retains recoverable prefixes in the beam for partial matching to turn into actual hits.

\textbf{User-semantic alignment directs tail opportunities to the right user.} To test whether user-semantic alignment merely adds random diversity, we fix popularity calibration and swap each user's query vector $\mathbf{q}_u$ with another user's. Both settings achieve similar tail exposure coverage (around 99\%). However, true user semantics consistently outperforms swapped semantics, raising tail R@200 from 20.63\% to 21.22\% on Industrial and from 18.41\% to 20.54\% on Office, and EffCov@200 from 23.48\% to 25.20\% and from 23.04\% to 25.03\%. This shows that user-semantic guidance retains tail prefixes matching the user's interest.

\textbf{Popularity calibration helps tail prefixes survive popular rivals.} Finally, we examine the decoding steps where all recoverable prefixes of a tail target are pushed out of the top-20 beam. Comparing the best dropped tail prefix against the irrelevant prefixes that stay in the beam, we find that 83.53\% to 95.45\% of these winning rivals on Industrial and 89.47\% to 91.21\% on Office have higher popularity priors. Applying single-step popularity calibration allows the dropped tail prefix to overtake 1.31\% to 3.86\% of these rivals, preventing popular branches from dominating the beam.

Together, these analyses clarify how each mechanism in ReSID contributes across the pipeline: popularity calibration corrects high-prior competition during pruning, user-semantic alignment retains tail prefixes matching user interests, and partial matching works with guided search to turn surviving local tokens into long-tail hits.

\subsection{Hyper-parameter Analysis}
\label{sec:exp-sens}

Figure~\ref{fig:sensitivity} investigates how the popularity-calibration
strength $\lambda$, the user-semantic weight $\beta$
and the mismatch tolerance $r$ affect performance, varying one hyper-parameter at a time while fixing the others at their
default values.
\sysname{} is insensitive to $\lambda$ and $\beta$: for
$\lambda\in[0.3,0.7]$ and $\beta\in[0.2,0.6]$, overall R@200 stays within
36.52--37.16 and long-tail R@200 within 19.79--20.67. A larger $\lambda$
gradually shifts the balance toward the tail, raising long-tail N@200 from
0.0451 to 0.0477 and long-tail EffCov@200 from 24.13\% to 25.29\%, while
overall R@200 decreases only marginally from 37.16 to 36.52. A larger $\beta$
mainly improves R@50, from 27.60 to 28.46 on all requests and from 8.55 to
9.11 on long-tail requests. The mismatch tolerance $r$ has the largest effect: long-tail R@200 rises
from 7.28 at $r{=}0$ to 18.45 at $r{=}2$ and 20.35 at $r{=}3$, and saturates
afterwards, as $r{=}4$ yields the same results as $r{=}3$. These results indicate that \sysname{} does not rely on careful
tuning, which is consistent with our use of a single set of guidance and
matching hyper-parameters on all datasets.


\subsection{Efficiency Analysis}
\label{sec:exp-eff}

\begin{figure}[!htbp]
  \centering
  \includegraphics[width=\columnwidth]{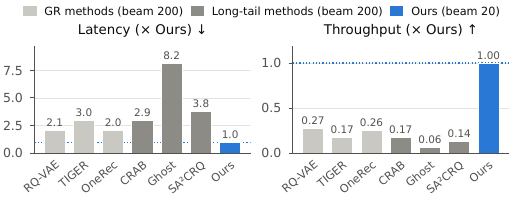}
  \caption{Serving cost of each method relative to \sysname{}, with latency and throughput.}
  \Description{Two grouped bar charts relative to ReSID. Left: latency multipliers of 2.1 for RQ-VAE, 3.0 for TIGER, 2.0 for OneRec, 2.9 for CRAB, 8.2 for Ghost, 3.8 for SA$^2$CRQ and 1.0 for ReSID. Right: throughput multipliers of 0.27, 0.17, 0.26, 0.17, 0.06, 0.14 and 1.00 for the same methods in the same order. ReSID has the lowest latency and the highest throughput.}
  \label{fig:efficiency}
\end{figure}

We measure end-to-end retrieval latency and throughput, covering history
encoding, generation, beam-search guidance, matching and ranking. All
methods return 200 unique items; Appendix~\ref{app:impl} details the protocol. As shown in
Figure~\ref{fig:efficiency}, where results are averaged over all datasets, the baselines incur
2.0--8.2$\times$ the latency of \sysname{} and achieve only
0.06--0.27$\times$ its throughput, with Ghost being the most expensive owing
to the longer SIDs of its tail items. Both popularity calibration and partial
matching are inference-time operations that add no model parameters, and
user-semantic alignment only adds a lightweight linear projection over the
frozen encoder; their overhead is outweighed by the savings from a narrower beam. 
On the In-house dataset, ReSID achieves a mean batch inference latency of 63.75 ms for batches of 512 requests on NVIDIA L20 GPUs, covering encoder computation and complete beam-search decoding and matching.

\subsection{Online A/B Testing}
\label{sec:exp-online}

We evaluate \sysname{} via a user-randomized online A/B test on a large-scale e-commerce advertising platform from July 1 to July 12, 2026, allocating 15\% randomized traffic to each bucket. The control group deploys TIGER-style generative retrieval system, while the treatment group replaces it with \sysname{}, keeping all other retrieval channels and downstream ranking unchanged. Long-tail items are identified following the same offline cumulative-click rule.
Compared to the control, \sysname{} improves advertising revenue, CTR, and GMV by 1.3\%, 1.1\%, and 0.8\%, respectively, while increasing long-tail exposure share by 4.0\% (all $p < 0.05$, two-sided test). With a mean online response time of 10.3~ms, \sysname{} has been fully deployed in production since July 2026 to serve main user traffic. 


\section{Conclusion}
In this paper, we study long-tail retrieval in generative recommendation from the perspective of item reachability across SID construction, decoding, and matching. Motivated by our stage-wise empirical findings, we propose \sysname{}, which pairs long parallel SIDs with popularity-debiased, user-semantic-guided beam search and partial SID matching to jointly preserve and recover tail candidates. Extensive offline and online evaluations verify the effectiveness and efficiency of \sysname{}.


\bibliographystyle{ACM-Reference-Format}
\begingroup
\catcode`\&=12
\bibliography{resid-exp}
\endgroup

\appendix


\section{Additional Reachability Analyses on Office}
\label{app:office-reachability}
We repeat the three controlled analyses on Office using seed 42 and beam 20. Head items have at least 12 TRAIN target occurrences; tail items have 1--6. Serial and parallel refer to offline SID construction; the online GR decoder is autoregressive in both cases. Recall is item-level. Effective coverage counts correctly retrieved distinct target items divided by distinct evaluation targets in the group; exposure coverage counts recommended distinct items divided by catalog items in the group.

\begin{figure}[h]
  \centering
  \includegraphics[width=\columnwidth]{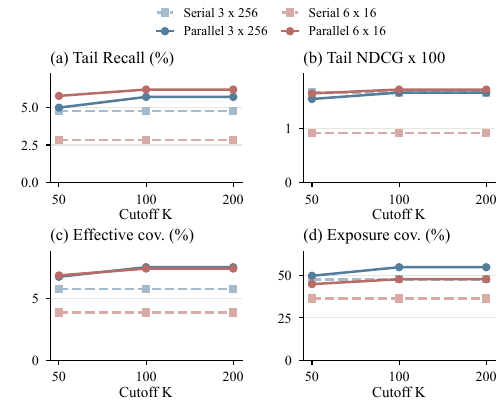}
  \caption{Office: construction comparison under exact matching. All panels use the same definitions and styles as Figure~\ref{fig:analysis-industrial-construction}.}
  \Description{Four panels show tail Recall, NDCG, effective coverage and exposure coverage at cutoffs fifty, one hundred and two hundred for four SID configurations.}
  \label{fig:analysis-office-construction}
\end{figure}

Parallel construction improves tail Recall at 100 and 200 and effective coverage under both code configurations (Figure~\ref{fig:analysis-office-construction}).

\begin{figure}[h]
  \centering
  \includegraphics[width=\columnwidth]{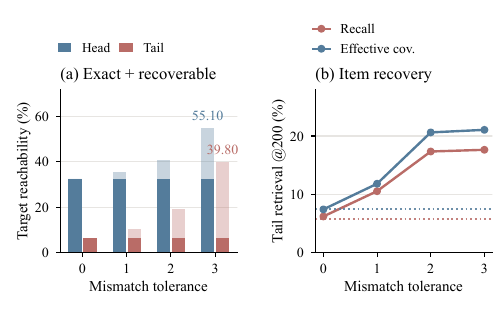}
  \caption{Office: retained local evidence and actual tail item recovery, with the same protocol as Figure~\ref{fig:analysis-industrial-recovery}. Dotted lines denote parallel $3\times256$ exact matching.}
  \Description{Cumulative target reachability and actual tail Recall and effective coverage at two hundred, as mismatch tolerance increases from zero to three.}
  \label{fig:analysis-office-recovery}
\end{figure}

Allowing three mismatches increases tail candidate reachability from 6.18\% to 39.80\%; actual tail Recall@200 rises from 6.18\% to 17.63\%, and effective coverage from 7.39\% to 21.06\% (Figure~\ref{fig:analysis-office-recovery}).

\begin{figure}[h]
  \centering
  \includegraphics[width=\columnwidth]{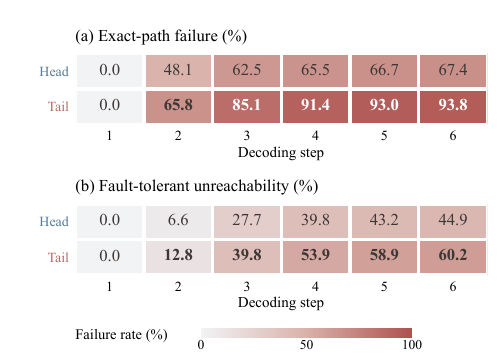}
  \caption{Office: early beam-search failure (parallel $6\times16$, beam 20). (a) Exact-path failure: the target prefix has left the beam. (b) Fault-tolerant unreachability: no retained prefix admits a catalog-valid completion within three mismatches of the target. Darker cells indicate higher failure. Final exact failure is 93.82\% for tail versus 67.42\% for head; 53.87\% of tail targets are already unrecoverable by step 4.}
  \Description{Two annotated heatmaps show head and tail cumulative failures across all six decoding steps. The upper heatmap measures exact prefix loss; the lower measures loss of all catalog-valid completions within three mismatches. Tail failure is substantially higher and occurs early.}
  \label{fig:analysis-office-boundary}
\end{figure}

Figure~\ref{fig:analysis-office-boundary} shows that post-decoding matching cannot recover the tail targets whose entire legal tolerance neighborhood has already been pruned.

\begin{table*}[t]
\caption{Performance comparison at small cutoffs, with Recall and NDCG at $K\in\{5,10,20\}$ and coverage at $K{=}100$. Marking follows Table~\ref{tab:main}.}
\label{tab:smallk}
\centering\scriptsize
\setlength{\tabcolsep}{6.55pt}
\begin{tabular}{@{}llcccccccccccccccc@{}}
\toprule
\multirow{2}{*}{Dataset} & \multirow{2}{*}{Method} & \multicolumn{2}{c}{R@5} & \multicolumn{2}{c}{R@10} & \multicolumn{2}{c}{R@20} & \multicolumn{2}{c}{N@5} & \multicolumn{2}{c}{N@10} & \multicolumn{2}{c}{N@20} & \multicolumn{2}{c}{EffCov@100} & \multicolumn{2}{c}{ExpCov@100} \\
\cmidrule(lr){3-4} \cmidrule(lr){5-6} \cmidrule(lr){7-8} \cmidrule(lr){9-10} \cmidrule(lr){11-12} \cmidrule(lr){13-14} \cmidrule(lr){15-16} \cmidrule(lr){17-18}
 & & All & Tail & All & Tail & All & Tail & All & Tail & All & Tail & All & Tail & All & Tail & All & Tail \\
\midrule
\multirow{7}{*}{Industrial} & RQ-VAE & 8.67 & 1.13 & 10.54 & 1.49 & 11.93 & 1.72 & 0.0662 & \underline{0.0067} & 0.0724 & 0.0079 & 0.0759 & 0.0084 & 17.71 & 4.47 & 75.37 & 57.27 \\
 & TIGER & 10.10 & 0.12 & 12.40 & 0.18 & 14.93 & 0.42 & \textbf{0.0840} & 0.0008 & \textbf{0.0914} & 0.0011 & \textbf{0.0978} & 0.0017 & 18.62 & 3.21 & 77.56 & 61.50 \\
 & OneRec & \textbf{10.70} & \underline{1.19} & \underline{13.08} & 2.20 & \textbf{16.72} & 4.34 & 0.0779 & 0.0060 & 0.0855 & 0.0092 & 0.0947 & 0.0146 & 22.64 & 10.08 & 77.40 & 62.06 \\
\cmidrule(l){2-18}
 & CRAB & 9.84 & 0.12 & 12.05 & 0.24 & 14.89 & 0.36 & \underline{0.0823} & 0.0006 & \underline{0.0893} & 0.0010 & \underline{0.0964} & 0.0013 & 18.97 & 2.98 & 80.01 & 66.20 \\
 & Ghost & 9.62 & 1.13 & 11.96 & \underline{2.50} & 14.82 & \underline{4.46} & 0.0800 & 0.0062 & 0.0875 & \underline{0.0105} & 0.0947 & \underline{0.0154} & \underline{23.27} & \underline{13.86} & \underline{92.73} & \underline{88.95} \\
 & SA$^2$CRQ & 7.81 & 0.65 & 9.66 & 1.01 & 11.96 & 1.49 & 0.0599 & 0.0040 & 0.0659 & 0.0052 & 0.0716 & 0.0065 & 18.40 & 4.24 & 37.36 & 18.31 \\
\cmidrule(l){2-18}
\rowcolor{oursrow} & \textbf{\sysname{}} & \underline{10.26} & \textbf{2.38} & \textbf{13.50} & \textbf{3.45} & \underline{16.02} & \textbf{5.11} & 0.0737 & \textbf{0.0142} & 0.0844 & \textbf{0.0176} & 0.0908 & \textbf{0.0218} & \textbf{26.02} & \textbf{16.95} & \textbf{93.35} & \textbf{90.21} \\
\addlinespace
\multirow{7}{*}{Office} & RQ-VAE & 11.39 & 0.43 & 13.44 & 0.78 & 16.24 & 1.14 & 0.0942 & 0.0028 & 0.1008 & 0.0040 & 0.1078 & 0.0049 & 21.60 & 3.75 & 82.10 & 68.53 \\
 & TIGER & \underline{12.08} & 0.57 & 13.89 & 0.85 & 16.44 & 1.14 & \textbf{0.1006} & 0.0040 & \textbf{0.1064} & 0.0049 & \underline{0.1128} & 0.0056 & 21.95 & 2.76 & 83.93 & 71.57 \\
 & OneRec & \textbf{12.31} & \textbf{1.71} & \textbf{14.39} & 2.06 & \textbf{17.76} & 2.20 & 0.0977 & \textbf{0.0121} & 0.1043 & \textbf{0.0132} & \textbf{0.1129} & 0.0135 & 24.41 & 6.28 & 81.01 & 66.77 \\
\cmidrule(l){2-18}
 & CRAB & 11.43 & 0.50 & 13.40 & 0.92 & 15.89 & 1.14 & \underline{0.0980} & 0.0042 & 0.1043 & 0.0055 & 0.1105 & 0.0061 & 22.52 & 3.53 & 85.08 & 74.13 \\
 & Ghost & 11.76 & \underline{1.49} & 13.01 & \underline{2.35} & 14.67 & \underline{3.70} & \textbf{0.1006} & \underline{0.0091} & \underline{0.1047} & 0.0118 & 0.1088 & \underline{0.0152} & \underline{25.38} & \underline{11.58} & \underline{94.62} & \underline{91.15} \\
 & SA$^2$CRQ & 10.62 & 0.28 & 12.68 & 0.50 & 15.82 & 0.71 & 0.0854 & 0.0016 & 0.0921 & 0.0023 & 0.1000 & 0.0028 & 18.42 & 2.43 & 35.62 & 17.23 \\
\cmidrule(l){2-18}
\rowcolor{oursrow} & \textbf{\sysname{}} & 11.04 & 1.28 & \underline{13.91} & \textbf{2.84} & \underline{17.00} & \textbf{4.48} & 0.0892 & 0.0073 & 0.0985 & \underline{0.0122} & 0.1064 & \textbf{0.0164} & \textbf{28.35} & \textbf{16.98} & \textbf{95.26} & \textbf{92.37} \\
\addlinespace
\multirow{7}{*}{In-house} & RQ-VAE & 14.28 & 4.13 & 18.28 & 5.26 & 22.53 & 6.51 & 0.1062 & 0.0310 & 0.1191 & 0.0346 & 0.1299 & 0.0378 & 35.91 & 9.98 & 44.04 & 36.14 \\
 & TIGER & \underline{14.52} & 3.60 & \underline{18.48} & 4.54 & \underline{22.73} & 5.63 & \underline{0.1083} & 0.0271 & \underline{0.1211} & 0.0302 & \underline{0.1318} & 0.0329 & \underline{36.03} & 8.59 & 45.01 & 34.07 \\
 & OneRec & 13.85 & 4.00 & 17.83 & 5.13 & 22.13 & 6.40 & 0.1024 & 0.0298 & 0.1153 & 0.0335 & 0.1261 & 0.0367 & 35.63 & 9.88 & 44.88 & 33.88 \\
\cmidrule(l){2-18}
 & CRAB & 11.33 & 2.56 & 14.73 & 3.28 & 18.48 & 4.10 & 0.0827 & 0.0191 & 0.0937 & 0.0214 & 0.1032 & 0.0235 & 30.73 & 6.27 & 44.03 & 23.45 \\
 & Ghost & 10.76 & \underline{6.51} & 14.12 & \underline{8.43} & 17.93 & \underline{10.65} & 0.0781 & \underline{0.0481} & 0.0889 & \underline{0.0543} & 0.0985 & \underline{0.0599} & 30.37 & \underline{16.51} & 43.85 & \underline{41.01} \\
 & SA$^2$CRQ & 12.24 & 5.05 & 15.98 & 6.47 & 20.11 & 8.11 & 0.0892 & 0.0375 & 0.1012 & 0.0421 & 0.1117 & 0.0463 & 33.69 & 12.51 & \underline{48.35} & 20.56 \\
\cmidrule(l){2-18}
\rowcolor{oursrow} & \textbf{\sysname{}} & \textbf{17.96} & \textbf{13.02} & \textbf{22.88} & \textbf{16.62} & \textbf{28.18} & \textbf{20.65} & \textbf{0.1332} & \textbf{0.0978} & \textbf{0.1492} & \textbf{0.1093} & \textbf{0.1625} & \textbf{0.1195} & \textbf{48.64} & \textbf{33.73} & \textbf{82.56} & \textbf{66.54} \\
\bottomrule
\end{tabular}
\end{table*}

\section{Experimental Details}
\label{app:details}

\subsection{Dataset Statistics}
\label{app:datasets}

\begin{table}[h!]
\caption{Dataset statistics and popularity distribution. Item shares are relative to the catalog and interaction shares to all training interactions.}
\label{tab:stats}
\centering
\small
\setlength{\tabcolsep}{4.5pt}
\begin{tabular}{lrrrrrr}
\toprule
\multirow{2}{*}{Dataset} & \multirow{2}{*}{Items} & \multirow{2}{*}{Inter.} & \multicolumn{2}{c}{Long-tail (\%)} & \multicolumn{2}{c}{Head (\%)} \\
\cmidrule(lr){4-5}\cmidrule(lr){6-7}
 & & & Items & Inter. & Items & Inter. \\
\midrule
Industrial & 3,686 & 36,259 & 53.77 & 22.44 & 20.56 & 56.46 \\
Office     & 3,459 & 38,924 & 54.21 & 20.20 & 22.67 & 62.54 \\
\bottomrule
\end{tabular}
\end{table}

Table 4 reports the public dataset statistics, where long-tail items exceed half the catalog but receive only 20--22\% of training interactions. For In-house (grouped by historical frequency where head items cover 80\% of interactions), long-tail items form 88.49\% of the 4,171,776 catalog items and account for 37.58\% of the 1,328,179 evaluation requests (spanning 662,788 distinct target items across all target sets).

\subsection{Baselines}
\label{app:baselines}

\begin{itemize}
  \item \textbf{RQ-VAE}~\cite{9879532} residually quantizes item
        content embeddings into three codes without any disambiguation code,
        so that items colliding on the same SID remain indistinguishable.
  \item \textbf{TIGER}~\cite{3666122.3666574} appends a collision code to the
        three-level RQ-VAE SID so that colliding items become distinguishable.
  \item \textbf{OneRec}~\cite{deng2025onerecunifyingretrieverank} builds SIDs with
        capacity-balanced residual $k$-means, bounding the number of items
        assigned to each code.
  \item \textbf{CRAB}~\cite{fan2026crabcodebookrebalancingbias} splits over-popular codes and combines
        a tree constraint with adaptation training to relieve the competition
        under head codes.
  \item \textbf{Ghost}~\cite{yin2026echoesfilterbubblediagnosing} lets tail items inherit head
        prefixes with longer tail-specific suffixes and adds tail
        unlikelihood training.
  \item \textbf{SA$^2$CRQ}~\cite{10.1145/3805712.3808477} combines adaptive code length
        with head-anchored codebook extension for tail items to address both
        collisions and tail fragmentation.
\end{itemize}

\subsection{Implementation and Efficiency Protocol}\label{app:impl}
\sysname{} and all baselines use the random seed 42; checkpoints are
selected on the validation split and ground truth is never fed at inference. We use multi-positive retrieval evaluation on In-house and leave-one-out evaluation with one held-out target per request on the public datasets, with all group metrics (including EffCov and ExpCov) computed over the requests within each respective group (All or Tail).
Ablation variants reuse the weights of the full model; the variant with
sequential SIDs reuses the GR trained on sequential $6{\times}16$ SIDs and
adapts the user-semantic projection with the same 8-epoch protocol, without
any new hyperparameter search. 
For the efficiency analysis on the public benchmarks (Industrial and Office),
all methods run on an Apple M4 CPU in FP32 with identical thread settings,
where \sysname{} decodes with a beam width of 20 and baselines use a beam
width of 200; each dataset uses the same 64 requests repeated for three
rounds, and every method returns 200 unique items. Timing covers history
encoding, generation, beam-search guidance, partial matching and ranking. Latency is the
mean per-request time and throughput the number of requests
per second.
On In-house (beam 100 for ReSID vs.\ 300 for baselines; NVIDIA L20 in BF16, batch size 512, history length 128), we approximate user-semantic prefix potentials via dot products with offline-precomputed normalized prefix centroids, and perform partial matching via prebuilt position--token inverted lists to accumulate token matches, filter candidates within tolerance, and rank items. End-to-end CUDA-synchronized latency (encoding, guided decoding, inverted-index matching, and ranking) averages 63.75~ms/batch, while the 10.3~ms online latency reflects single-request serving in our C++/CUDA engine.

\section{Additional Results}
\label{app:results}

\subsection{Performance at Small Cutoffs}
\label{app:smallk}

Table~\ref{tab:smallk} reports Recall and NDCG at $K\in\{5,10,20\}$ and the
coverage metrics at $K{=}100$. On Industrial, \sysname{} remains the best on
all long-tail metrics, and on In-house it is the best on every metric for
both all and long-tail requests. On Office, it is the best on long-tail R@10,
R@20, N@20 and both coverage metrics, while OneRec is ahead on long-tail R@5,
N@5 and N@10. Both coverage metrics of \sysname{} are the highest on all
requests and on the tail on all three datasets.

\subsection{Detailed Results for Fine-Grained Mechanism Analysis}
\label{app:mech}

We report the complete numerical results for the four mechanism analyses in Section~\ref{sec:mech}.

\paragraph{Contribution of Partial Matching.} 
Table~\ref{tab:mech-partial} breaks down retrieved tail targets into \emph{Total Hits} and \emph{Partial-Only Hits} (hits with no exact match in the decoded beams). At $K=200$, partial matching accounts for over half of all tail hits on both datasets.

\begin{table}[h!]
\caption{Breakdown of long-tail hits in \sysname{} across cutoffs.}
\label{tab:mech-partial}
\centering
\small
\begin{tabular}{lcccc}
\toprule
Dataset & $K$ & Total Hits & Partial-Only Hits & Partial Share \\
\midrule
Industrial & 50  & 155 & 20  & 12.90\% \\
Industrial & 100 & 236 & 71  & 30.08\% \\
Industrial & 200 & 357 & 183 & \textbf{51.26\%} \\
\addlinespace
Office     & 50  & 114 & 18  & 15.79\% \\
Office     & 100 & 192 & 94  & 48.96\% \\
Office     & 200 & 289 & 191 & \textbf{66.09\%} \\
\bottomrule
\end{tabular}
\end{table}

\paragraph{Joint Preservation and Recovery.} 
Table~\ref{tab:mech-synergy} examines tail targets unreachable within three mismatches under standard beam search yet retrieved by \sysname{} at $K=200$. After guided search preserves their recoverable prefixes in the beam, around 90\% of these newly rescued targets are ultimately retrieved through partial matching.

\begin{table}[h!]
\caption{Breakdown of newly rescued tail hits at $K=200$.}
\label{tab:mech-synergy}
\centering
\small
\begin{tabular}{lccc}
\toprule
Dataset & Rescued Hits & Exact Hits & Partial Hits (Share) \\
\midrule
Industrial & 10 & 1 & \textbf{9 (90.00\%)} \\
Office     & 9  & 1 & \textbf{8 (88.89\%)} \\
\bottomrule
\end{tabular}
\end{table}

\paragraph{Effect of User-Semantic Alignment.} 
Table~\ref{tab:mech-semantic} compares guiding the beam with each user's own query (\emph{True}) versus another user's query (\emph{Swapped}) under fixed popularity calibration. While both expose over 99\% of tail items, true user semantics consistently improves tail Recall and effective coverage.

\begin{table}[h!]
\caption{Comparison between swapped and true user-semantic guidance under fixed popularity calibration.}
\label{tab:mech-semantic}
\centering
\small
\setlength{\tabcolsep}{3.5pt}
\begin{tabular}{llccccc}
\toprule
Dataset & Query & R@50 & R@100 & R@200 & EffCov & ExpCov \\
\midrule
Industrial & Swapped & 8.68\% & 13.91\% & 20.63\% & 23.48\% & 99.04\% \\
Industrial & \textbf{True} & \textbf{9.22\%} & \textbf{14.03\%} & \textbf{21.22\%} & \textbf{25.20\%} & \textbf{99.09\%} \\
\addlinespace
Office     & Swapped & 6.89\% & 12.44\% & 18.41\% & 23.04\% & 99.15\% \\
Office     & \textbf{True} & \textbf{8.10\%} & \textbf{13.65\%} & \textbf{20.54\%} & \textbf{25.03\%} & \textbf{99.31\%} \\
\bottomrule
\end{tabular}
\end{table}

\paragraph{Effect of Popularity Calibration.} 
Table~\ref{tab:mech-pop} pairs the best dropped tail prefix against irrelevant in-beam prefixes (\emph{Competing Pairs}) at steps 4--6. Over 83\%--95\% of winning rivals have higher popularity priors (\emph{Higher-Prior Rivals}), and single-step calibration enables the tail prefix to overtake 1.31\%--3.86\% of them (\emph{Overtaken}).

\begin{table}[h!]
\caption{Pairwise competition between dropped tail prefixes and in-beam rivals at decoding steps 4--6.}
\label{tab:mech-pop}
\centering
\small
\setlength{\tabcolsep}{4pt}
\begin{tabular}{lcccc}
\toprule
Dataset & Step & Competing Pairs & Higher-Prior Rivals & Overtaken \\
\midrule
Industrial & 4 & 5,020 & \textbf{83.53\%} & 2.17\% \\
Industrial & 5 & 2,060 & \textbf{91.41\%} & 1.31\% \\
Industrial & 6 & 660   & \textbf{95.45\%} & 2.73\% \\
\addlinespace
Office     & 4 & 3,960 & \textbf{91.14\%} & 2.32\% \\
Office     & 5 & 1,400 & \textbf{91.21\%} & 3.86\% \\
Office     & 6 & 380   & \textbf{89.47\%} & 2.63\% \\
\bottomrule
\end{tabular}
\end{table}

\newtheorem{assumption}[theorem]{Assumption}
\section{Theoretical Analysis of Bias Propagation in SID Construction}
\label{app:theory}
In this section, we formally explain why sequential SIDs accumulate popularity bias across layers while parallel SIDs isolate each position. Consider a fixed tail item vector $\mathbf{z}_i$, fixed subspace projections $P_\ell$, and nearest-centroid quantization. We compare an ideal unbiased quantizer against a head-biased quantizer (denoted by primes).

\begin{definition}[Quantization displacement]
\label{def:quantization-drift}
In sequential quantization, let $\mathbf{r}_1=\mathbf{z}_i$, $c_\ell=Q_\ell(\mathbf{r}_\ell)$, $\mathbf{u}_\ell=\mathbf{e}_\ell(c_\ell)$, and $\mathbf{r}_{\ell+1}=\mathbf{r}_\ell-\mathbf{u}_\ell$ be the ideal residual trajectory, and let $\mathbf{r}'_\ell, \mathbf{u}'_\ell$ be the perturbed trajectory under a head-biased quantizer starting from $\mathbf{r}'_1=\mathbf{z}_i$. We define the single-step codeword shift $\boldsymbol{\delta}_\ell$ and the cumulative residual drift $\Delta\mathbf{r}_\ell$ at layer $\ell$ as
\begin{equation}
\boldsymbol{\delta}_\ell=\mathbf{u}'_\ell-\mathbf{u}_\ell,
\qquad
\Delta\mathbf{r}_\ell=\mathbf{r}'_\ell-\mathbf{r}_\ell.
\label{eq:drift-definition}
\end{equation}
Given a unit vector $\mathbf{d}$ pointing toward head-dominated regions, we define the accumulated directional bias as $B_\ell=-\mathbf{d}^\top\Delta\mathbf{r}_\ell$.
\end{definition}

\begin{assumption}[Aligned head-directed shifts]
\label{ass:head-directed}
At each preceding layer $j<\ell$, the codeword shift is positively aligned with the head direction, satisfying $\mathbf{d}^\top\boldsymbol{\delta}_j\ge b_j\ge0$.
\end{assumption}

We first show that in sequential quantization, codeword shifts from early layers add up and push later residuals progressively further from their ideal values.

\begin{proposition}[Sequential accumulation]
\label{prop:serial-accumulation}
Sequential residual quantization satisfies
\begin{equation}
\Delta\mathbf{r}_\ell=-\sum_{j<\ell}\boldsymbol{\delta}_j.
\label{eq:residual-bias}
\end{equation}
Under Assumption~\ref{ass:head-directed}, its directional bias obeys $B_\ell\ge\sum_{j<\ell}b_j$. In particular, if $b_j\ge b>0$, then $B_\ell\ge(\ell-1)b$.
\end{proposition}
\begin{proof}
Subtracting the ideal residual update from the biased update gives
\begin{equation}
\begin{aligned}
\Delta\mathbf{r}_{\ell+1}
&=(\mathbf{r}'_\ell-\mathbf{u}'_\ell)-(\mathbf{r}_\ell-\mathbf{u}_\ell)\\
&=\Delta\mathbf{r}_\ell-\boldsymbol{\delta}_\ell.
\end{aligned}
\end{equation}
Since $\Delta\mathbf{r}_1=\mathbf{0}$, unrolling this recursion yields Eq.~\eqref{eq:residual-bias}. Projecting onto $\mathbf{d}$ gives
\begin{equation}
B_\ell=\sum_{j<\ell}\mathbf{d}^\top\boldsymbol{\delta}_j
\ge\sum_{j<\ell}b_j\ge(\ell-1)b,
\end{equation}
which grows linearly with depth $\ell$.
\end{proof}

Next, we show that once this cumulative residual drift $\Delta\mathbf{r}_\ell$ exceeds a threshold, it flips the nearest-centroid assignment at downstream layers.

\begin{lemma}[Assignment sensitivity]
\label{lem:assignment-margin}
At a downstream layer, let $\mathbf{a}$ be the ideal centroid for residual $\mathbf{r}$ and $\mathbf{b}$ be a competing centroid, with positive distance margin $m_{ab}(\mathbf{r})=\|\mathbf{r}-\mathbf{b}\|^2-\|\mathbf{r}-\mathbf{a}\|^2>0$. Under a residual drift $\Delta\mathbf{r}$, the margin changes by
\begin{equation}
\begin{aligned}
m_{ab}(\mathbf{r}+\Delta\mathbf{r})-m_{ab}(\mathbf{r})
=2\Delta\mathbf{r}^\top(\mathbf{a}-\mathbf{b}).
\end{aligned}
\label{eq:margin-shift}
\end{equation}
If $2\Delta\mathbf{r}^\top(\mathbf{a}-\mathbf{b}) < -m_{ab}(\mathbf{r})$, the competing centroid $\mathbf{b}$ becomes closer than $\mathbf{a}$ and flips the assigned token.
\end{lemma}
\begin{proof}
Expanding both squared distances cancels the common $\|\mathbf{r}\|^2$ and $\|\Delta\mathbf{r}\|^2$ terms and leaves Eq.~\eqref{eq:margin-shift}. When the right-hand side falls below $-m_{ab}(\mathbf{r})$, we have $m_{ab}(\mathbf{r}+\Delta\mathbf{r})<0$, which reverses the nearest-centroid order.
\end{proof}

Finally, we show that parallel quantization avoids this cascading error because each position quantizes its own subspace independently.

\begin{proposition}[Parallel isolation]
\label{prop:parallel-isolation}
In parallel construction $c_\ell=Q_\ell(P_\ell\mathbf{z}_i)$, perturbing the quantizer $Q_j$ at position $j$ leaves all other tokens $c_\ell$ ($\ell\ne j$) unchanged.
\end{proposition}
\begin{proof}
For every $\ell\ne j$, both the subspace input $P_\ell\mathbf{z}_i$ and the quantizer $Q_\ell$ remain unchanged, so
\begin{equation}
c'_\ell=Q'_\ell(P_\ell\mathbf{z}_i)
=Q_\ell(P_\ell\mathbf{z}_i)=c_\ell.
\end{equation}
Any quantization shift at position $j$ stays within its own subspace and never propagates to other positions.
\end{proof}

Consequently, parallel SIDs preserve clean token evidence at remaining positions even when a local shift occurs. Consistent with this analysis, switching from sequential to parallel $6\times16$ SIDs on Industrial reduces the tail full-SID negative log-likelihood (NLL) from 11.8923 to 10.6566 and narrows the head-to-tail NLL gap by 17.14\%.

\clearpage
\end{document}